\documentclass[preprint,showpacs,preprintnumbers,amsmath,amssymb]{elsarticle}
\usepackage{ae,graphicx}
\usepackage{amssymb}
\input{epsf}

\journal{Physica A}
\begin{document}

\title{Dynamical behavior of the Niedermayer algorithm applied to Potts models}

\author[icex,inct]{D. Girardi \corref{cor1}} 
\ead{girardi@if.uff.br}
\author[icex,inct,uff]{T. J. P. Penna}
\author[ufsc]{N. S. Branco}

\address[icex]{Instituto de Ci\^{e}ncias Exatas, Universidade Federal Fluminense - Rua Des. Ellis Hermydio Figueira, 783,Volta Redonda - RJ - 27.213-350- Brazil - Phone:(+55)24-30768944}
\address[inct]{National Institute of Science and Technology for Complex System - Brazil}

\address[uff]{Instituto de F\'isica, Universidade Federal Fluminense, Niter\'oi, RJ, Brazil}
\address[ufsc]{Departamento de F\'{\i}sica, Universidade Federal de Santa Catarina, 88040-900, Florian\'opolis, SC, Brazil}

\date{\today}

\begin{abstract}

         In this work we make a numerical study of the dynamic universality class of the
Niedermayer algorithm applied to the two-dimensional Potts model with 2, 3, and 4 states.
This algorithm updates clusters of spins and has a free parameter, $E_0$, which controls the size of these
clusters, such that $E_0=1$ is the Metropolis algorithm and $E_0=0$ regains the
Wolff algorithm, for the Potts model. For $-1<E_0<0$, only clusters of equal spins can be formed: we show that the mean size
of the clusters of (possibly) turned spins initially grows with the linear size of the lattice,
$L$, but eventually saturates at a given lattice size $\widetilde{L}$, which
depends on $E_0$. For $L \geq \widetilde{L}$, the Niedermayer algorithm is in the same dynamic universality class of
the Metropolis one, i.e, they have the same dynamic exponent. For $E_0>0$, spins
in different states may be added to the cluster but the dynamic behavior is less efficient than for the Wolff
algorithm ($E_0=0$).
Therefore, our results show that the Wolff algorithm is the best choice for Potts models, when compared
to the Niedermayer's generalization.

\end{abstract}

%\noindent{ Numerical methods; dynamic behavior; Cluster algorithms; Potts model}
%\pacs{ 07.05.Tp; 05.10.Ln; 05.10-a}

\maketitle

%\newpage

\section{Introduction} \label{sec:introduction}

   Numerical simulations of physical systems have been used for decades (for a review of numerical
methods in statistical physics, see Refs. \cite{landau} and \cite{newman}). In recent
years, however, this method has received a renewed interest, due to the increase
in computational power and, more important, due to the introduction of new algorithms.
The developments in the algorithms have the goal to allow for more efficient simulations,
in many different directions: the calculation of the density of states using flat histograms,
which allows for obtaining information
at any temperature from one single simulation \cite{wang};
the use of bitwise operations and storage, which increases by a great amount the speed of
the simulation and saves memory \cite{penna1990}; and the
introduction of cluster algorithms, which updates collections of spins, decreasing
the autocorrelation time and reducing critical slowing down \cite{landau,newman,
swendsen, wolff}.

      The Metropolis algorithm \cite{metropolis}, for example, which had been the main choice of
algorithm for a long time, suffers from a severe critical slowing-down effect near critical points. 
This is in part due to the fact that it updates one spin each time (therefore, being in the
general category of single-spin algorithms). Critical slowing down is measured through the
dynamic critical exponent $z$, defined from the dependence of the autocorrelation time $\tau$
on the linear size of the simulated lattice, $L$, at the critical temperature $T_c$. This dependence
 is assumed to be in the form $\tau \sim L^z$. Therefore, smaller values of  $z$ lead to smaller
 autocorrelation times and more efficient algorithms, since more configurations can be used to
 calculate the necessary averages. The Metropolis algorithm, for example, when applied to the Ising model in two dimensions,
presents $z \sim 2.16$ \cite{nightingale}.

      One possible way to overcome the difficulty of critical slowing down is to design algorithms
which update clusters of spins (the so-called cluster algorithms), that may have a
much lower value of $z$: this is the case for the Swendsen-Wang \cite{swendsen}
and Wolff \cite{wolff} algorithms, for which $z$ may be zero 
for the two-dimensional Ising model \cite{baillie,coddington}. See also Ref. \cite{girardi} for another
possible dependence of the autocorrelation time on the lattice size $L$.
An alternative (and generalization) to these last two cluster algorithms, the
Niedermayer algorithm, was introduced some time ago \cite{niedermayer} but,
to the best of our knowledge, has only had his dynamic behavior
studied in detail for the Ising and $XY$ models \cite{girardi}. Our goal in this work is to
study in a systematic way the dynamic behavior of the Niedermayer algorithm, applied to Potts models in two dimensions,
for number of states $q=2$, $3$, and $4$, 
for some values of the free parameter $E_0$ (see below), in
order to determine the best choice of this parameter for these models. Note also that the critical temperature
for two-dimensional Potts models are exactly known \cite{wu},
which allows for a more precise determination of $z$.

         The Potts model is defined by the Hamiltonian \cite{wu}
\begin{equation}
     {\cal H} = -J \sum_{<i,j>} \delta_{S_i,S_j},  \label{hamiltonianapotts}
\end{equation}
where $S_i=1, 2, \ldots, q$, $\forall i$, the sum is over nearest neighbors on a lattice (in our case, a square lattice)
and $\delta_{S_i,S_j}$ is the Kronecker delta ($\delta_{S_i,S_j}=1$, if $S_i=S_j$, and $\delta_{S_i,S_j}=0$,
if $S_i \neq S_j$). We treat the ferromagnetic case in this work, i.e, $J>0$. In two dimensions, the phase transition
for this model is a continuous one for $q \leq 4$. In our study, we restrict ourselves to this interval.
 
	This work is organized as follows: in the next section we present
the Niedermayer algorithm for the Potts model and its relation to Metropolis' and Wolff's. In Section
\ref{sec:timeandexponent} we review some features connected to the autocorrelation
time and the dynamic exponent $z$, in Section \ref{sec:results} we present
and discuss our results, and in the last section we summarize the results.

\section{The Niedermayer algorithm} \label{sec:algorithm}

	The Niedermayer algorithm was introduced some time ago as a cluster algorithm,
motivated by the possibility to diminish the value of the dynamic
critical exponent. It may be seen as a generalization of the Wolff or the Swendsen-Wang algorithms.
The main idea is to build a cluster of spins and accept its updating
as a single entity,  with a parameter $E_0$ which controls the nature and size of the cluster and its
acceptance ratio, as explained below. In this work,
we have chosen to build the clusters according to the Wolff algorithm (they
can be constructed according to the Swendsen-Wang rule but
the results will not differ qualitatively in two
dimensions and in higher dimensions Wolff algorithm is superior
to Swendsen-Wang's). 

     In the Niedermayer algorithm, a spin in the lattice, which we will call the {\it seed} spin, is randomly chosen, being the
first spin of the cluster. First-neighbors of this spin may be
considered part of the cluster, with a probability
	\begin{equation}
	P_{add} (E_{ij}) = \left\{ \begin{array}{lcl}
	                     1-e^{K \left( E_{ij}-E_0 \right) } & ,  & \mbox{if} \;\; E_{ij}<E_0 \\
                             0 & ,                                      & \mbox{otherwise}
	                  \end{array},
                 \right.  \label{eq:Padd}
	\end{equation}
where $K = J/kT$, $T$ is the temperature, $J$ is the exchange constant,
and $E_{ij}$ is the energy between
nearest-neighbor spins in unities of $J$ (i.e, $E_{ij} =- \delta_{ s_i s_j}$). First-neighbors of added spins may be
added to the cluster, according to the probability given above. 
Each spin has more than one chance to be part of the cluster, since it
may have more than one first-neighbor in it. When no
more spins can be added, all spins in the cluster have their state changed to a new state
with an acceptance ratio $A$. Assuming that, at the frontier of the cluster
there are $m$ bonds linking spins in the same state \textit{in the old configuration} 
and $n$ bonds linking spins in the same state
\textit{in the new configuration} (there are also $p$ bonds in the border of the cluster which
are in different states in the old and in the new configurations, but they cancel
out in the expression below), $A$ satisfies:
\begin{equation}
   \frac{A(a\rightarrow b)}{A(b\rightarrow a)} = \left[ e^{ K}
         \left(  \frac{1-P_{add}(-J)}{1-P_{add}(0)} \right) \right]^{n-m}, \label{eq:acceptanceratio}
\end{equation}
where $a\rightarrow b$ represents the possible updating process, 
from the ``old'' ($a$) to the ``new'' ($b$) state, which differ from the
flipping of all spins in the cluster, and $b\rightarrow a$ represents the
opposite move. This expression ensures that detailed balance is satisfied \cite{newman}.

	One has to consider three different intervals for $E_0$:
\begin{itemize}
\item [(i)] for $-1 \leq E_0 < 0$, only spins in the same state as the seed
may be added to the cluster, with probability $P_{add} = 1 - e^{-K(1+E_0)}$.
The new state of the cluster is randomly chosen between the remaining $q-1$ states.
The acceptance ratio (Eq. \ref{eq:acceptanceratio}) cannot be chosen to be one
always and is given by $A = e^{-K E_0 (m-n)}$, if $n<m$ (i.e, if the energy
increases when the spins in the cluster are changed), or by $A=1$, if
$n>m$ (i.e, if the energy
decreases when the spins in the cluster are changed). If $E_0=-1$, we obtain 
the Metropolis algorithm, since only one-spin clusters are allowed
(according to Eq. (\ref{eq:Padd}), $P_{add}=0$ for $E_0=-1$) and the
acceptance ratio is $A = e^{-K \Delta E}$ for positive $\Delta E$ and $1$
otherwise, where $\Delta E = (m-n)$ is the difference in energy when the spin is
changed, in units of $J$;
\item [(ii)] for $E_0 = 0$, again only spins in the same state can take part of
the cluster, with probability $P_{add} = 1 - e^{-K}$. Now, the acceptance
ratio can be chosen to be $1$, i.e, the cluster of like spins is always
changed. Again, the new state of the cluster is randomly chosen between the remaining $q-1$ states.
This is the celebrated Wolff algorithm;
\item [(iii)] for $E_0 > 0$, spins in different states
may be part of the cluster. Consider a spin already in the cluster: the probability of adding one 
of its first-neighbors to the cluster is $P_{add} = 1 - e^{-K(1+E_0)}$ if they are in the
same state or $P_{add} = 1 - e^{-K E_0}$ otherwise.
The acceptance ratio is again always $1$. To change the state, for each cluster, we randomly choose a $\Delta q$ between $1$ and $q-1$ and perform a cyclic sum. Note
that for $E_0 \gg 0$ nearly all spins will be in the cluster and
the algorithm will be clearly inefficient (in fact, it will not be ergodic for
$E_0 \rightarrow \infty$). Therefore, we expect that, if the optimal choice of
$E_0$ is greater than $0$, it will not be much greater than this value.
\end{itemize}

     After constructing a cluster, possible updating it and calculating the relevant thermodynamic functions
for the new configuration, the whole process is repeated with a new seed spin. In this way, the Markov
chain of configurations is generated.

\section{Autocorrelation time and dynamic exponent} \label{sec:timeandexponent}

         To calculate the relevant averages from a numerical simulation, one has to 
 build a Markov chain of spin configurations and use data from uncorrelated configurations along this chain.
 Therefore, one important quantity is the autocorrelation time for a given quantity, say $\Phi(t)$,
 obtained from the autocorrelation function $\rho(t)$:
 \begin{eqnarray}
   \rho(t) & = & \int \left[ \Phi(t') - <\Phi> \right] \left[ \Phi(t'+t) - <\Phi> \right] dt' \nonumber \\ 
   & = & \int \left[ \Phi(t') \Phi(t'+t) - <\Phi>^2 \right] dt', \label{eq:autocorrelation}
\end{eqnarray}
Since time is a discrete quantity on Monte Carlo simulations, one has to discretize the above equation \cite{newman}:
\begin{eqnarray}
   \rho(t) & = & \frac{1}{t_{max}-t} \sum_{t'=0}^{t_{max}-t} \left[ \Phi(t') \Phi(t'+t) \right] - \nonumber \\ 
    & & \frac{1}{(t_{max}-t)^2} \sum_{t'=0}^{t_{max}-t}  \Phi(t') \times \sum_{t'=0}^{t_{max}-t} \Phi(t'+t)
     \label{eq:discrete}
\end{eqnarray}

   It is usually assumed that the autocorrelation function behaves, in its simplest form, as \cite{newman}:
\begin{equation}
   \rho(t) = A e^{-t/\tau}, \label{eq:functionoftime}
\end{equation}
This hypothesis has to be corroborated by data and, in some cases, more than one
exponential term is required \cite{wansleben}. One point worth mentioning is that the
autocorrelation function is not well behaved
for long times, due to bad statistics (this is evident from Eq. \ref{eq:discrete},
since few ``measurements'' are available for long times). Then, one has to
choose the region where the
straight line will be adjusted very carefully and it turns out that the
value of $\tau$ so obtained is strongly dependent on this choice. One other
possible way to measure $\tau$ is to integrate $\rho(t)$, assuming
a single exponential dependence on (past and forward) time, and obtain:
\begin{equation}
   \tau = \frac{1}{2} \int_{-\infty}^{\infty} \frac{\rho(t)}{\rho(0)} dt, \label{eq:tau}
\end{equation}
with:
\begin{equation}
   \rho(t) \equiv e^{-|t|/\tau}.
\end{equation}
   Eq. \ref{eq:tau}, when discretized, leads to \cite{salas}:
\begin{equation}
   \tau = \frac{1}{2} + \sum_{t=1}^{\infty} \frac{\rho(t)}{\rho(0)}. \label{eq:somadetau}
\end{equation}

   The sum in the previous equation cannot be carried out for large values of $t$,
 since bad statistics would lead to unreliable results. In order to truncate the sum at some
 point, we use a cutoff (see Ref.\cite{salas} and references
therein), defined as the value in time where the noise in the data is clearly greater 
than the signal itself. We then obtain a first estimate of $\tau$ using Eq. \ref{eq:somadetau} and
then make the integral of $\rho(t)/\rho(0)$
from the value of the cutoff to infinity. A criterion to accept the cutoff is that the value of this
integral is smaller than the statistical uncertainty in calculating $\tau$. 

    A note on the definition of ``time'' is worth stressing here. In the Metropolis algorithm,
time is measured in Monte Carlo steps ($MCS$); one $MCS$ is defined as
the attempt to flip $N$ spins, where $N$ is the number of spins in
the (finite) lattice being simulated (in our case, $N=L^2$, where
$L$ is the linear size of the lattice). For cluster algorithms, one unity of time is defined as 
the ``time'' taken to build and possibly change a cluster, $t$. A rescaling of the quantity is necessary, to
be able to compare the results for different values of $E_0$, namely:
\begin{equation}
   t_{MCS} = t \frac{<n>}{N},
\end{equation}
where $t_{MCS}$ is the time measured in $MCS$ and $<n>$ is the mean cluster 
size. Note that, for Metropolis, $<n>=1$ and $1$ $MCS$
is the ``time'' taken to try to flip $N$ spins, as usual. In this work, this rescaling has been
done and all times are expressed in $MCS$. 

  Our first attempt was to fit the autocorrelation time to the expected behavior, namely
$\tau \sim L^z$, in the critical region, where $z$ is the dynamic exponent. We have taken as
the chosen quantity to calculate $\tau$ the one which better adjusts to Eq. \ref{eq:functionoftime}. Usually,
different quantities lead to autocorrelation functions which behave differently as a function
of time. A typical example
is shown in Fig. \ref{fig:magnatizacaoeenergia}, where both the magnetization 
and the energy autocorrelation functions for the $q=3$ Potts model are depicted as functions of time, for the
Niedermayer
algorithm with $E_0=-0.25$ and linear sizes $L=16$ (main graph) and $L=128$ (inset).
The magnetization autocorrelation function presents an abrupt drop for small times and $L=16$. 
This shows that this function is not properly described
by a single exponential. On the other hand, the energy autocorrelation time
follows a straight line even for the smallest times. Therefore, we should calculate $\tau$ from the
latter, for $L=16$, using Eq. \ref{eq:somadetau}.
Note, however, that, when $L$ is increased, the picture changes and now  the magnetization
autocorrelation function is well described by a single exponential (for small and intermediate values of
time), as depicted in the inset of Fig. \ref{fig:magnatizacaoeenergia}. Whenever a
crossover like this is present, we measure the dynamic exponent from the behavior 
for large values of $L$ and for the function which is well described by a single exponential for this
range of $L$, using Eq. \ref{eq:somadetau}. The fact that, for intermediate values of $t$,  
the slopes of both curves in Fig. \ref{fig:magnatizacaoeenergia} (main graph and inset) appear to be the same, is
an indication that the autocorrelation times for both are the same. 
However, we have already
commented on the drawback of calculating $\tau$ from the slope of the autocorrelation function on a semi-log graph.
As final note, we would like to mention that we used helical boundary conditions and 20 independent runs (each with a
different seed for the pseudo random number generator) were made for each $E_0$ and $L$. For each seed, at least
$4 \times 10^6$ trial changes were made, in order to calculate the autocorrelation functions and their respective autocorrelation
times. The values we quote are the average of the values obtained for each seed of the pseudo random number generator and
the uncertainty in $\tau$ is the standard deviation of these $20$ values.

\section{Results and Discussion} \label{sec:results}

   We have simulated the case $E_0=-1$ (Metropolis algorithm) as a test to our code. 
The result is presented in Fig. \ref{fig:metropolis}  for the magnetization autocorrelation time and,
as we can see, all three cases have the same qualitative behavior. For $q=2$ and $3$ the 
dynamic exponent $z$ takes approximately the same value 
($z\simeq2.16$), while for $q=4$ it assumes a higher value, namely $z=2.21\pm0.02$. These values 
are consistent with those in the literature \cite{binder}.

	From now on we will not comment on $q=2$, since this case has been treated in Ref. 
\cite{girardi}. Our simulation for $E_0 = -0.75$ is presented in Fig. \ref{fig:tao-0,75}, where the magnetization and
energy autocorrelation times are depicted as functions of $L$. For $q=3$ and $L \geq 32$ and for $q=4$ and 
$L \geq 64$, the dynamic behavior (i.e., $z$) is the same as for the Metropolis algorithm.
In Figs. \ref{fig:tao-0,75}$b$ ($q=3$) and  \ref{fig:tao-0,75}$d$ 
($q=4$) we present the 
behavior of the average cluster size $<n>$ versus lattice size $L$: in both cases, for $L \geq \tilde{L} = 64$, 
$<n>$ is constant. As 
already discussed, the autocorrelation function which is well described by only one exponential presents the 
greatest autocorrelation time and for, $E_0=-0.75$, this happens for the magnetization's autocorrelation function.
The dynamic exponent is calculated for $L \geq \tilde{L}$ and we obtain $z=2.16 \pm 0.05$ and
$z= 2.18 \pm 0.09$ for $q=3$ and $q=4$, respectively. These results agree, within error bars, with the values 
quoted in the literature \cite{binder} and with our values obtained for $E_0=-1$.
	
	For $E_0=-0.25$, our result is depicted in Fig. \ref{fig:tao-0,25} for $q=3$ ( the result $q=4$ is qualitatively the same);
the behavior follows the same overall trend observed for $E_0=-0.75$, with a
different value of $\tilde{L}$. For $L<128$ the autocorrelation time for the energy is greater than for the magnetization and the average cluster size $<n>$ increases with $L$. For $L \geq \tilde{L} = 128$, the picture changes and the autocorrelation time for the magnetization is the greater one and $<n>$ is constant. The value of $z$ is consistent with the one for the Metropolis algorithm (again, $z$ is calculated for $L\geq \tilde{L}$).

        For other values of $-1\leq E_0 < 0 $ we observe the same qualitative behavior. There is always a $\tilde{L}$, such that,
for $L<\tilde{L}$, $<n>$ increases with $L$ and, for $L>\tilde{L}$, $<n>$ is constant and the magnetization's autocorrelation
time is the one well described by a single exponential. The exponent $z$, always calculated for $L \geq \tilde{L}$, is the
same as for the Metropolis algorithm. This can be linked to the constancy of $<n>$ in this interval: since $L$ increases and $<n>$
remains the same, the fraction $<n>/L^2$ decreases and the behavior is the same as a single-spin algorithm.

	We simulated the Wolff algorithm to perform another check of our algorithm and to compare with our results
for $E_0 \neq 0$. In Fig. \ref{fig:wolff} we present the autocorrelation time and $<n>$ versus $L$ for $q=3$ and $q=4$. To 
calculate $z$ we used a different approach here \cite{girardi1,picco}. 
We perform a power-law fitting using three consecutive lattice size 
(e.g $L=512$, $1024$, and $2048$) and call $L_{min}$ the smallest size. We then plot $z$ versus $1/L_{min}$,
as seen in  Fig. \ref{fig:picco}  \cite{girardi1,picco}. The power-law fitting for $q=3$ is a good fit for all lattice sizes
 and we obtain $z=0.55\pm0.02$, which
agrees with a previous estimate \cite{coddington2}. For  $q=4$, only for $L \geq 128$ we obtain a good fit, with $z=1.00\pm0.02$. 
This value is somewhat above a previous calculation \cite{li} for the Swendsen-Wang algorithm. If we restrict our data to
$L \leq 256$, as in the previous reference, we obtain $z=0.94 \pm 0.01$, which just overlaps with the result
of Ref. \cite{li} (namely, $z = 0.89 \pm 0.05$).
We plot the data for $q=2$ for comparison with our results for $q=3$ and $4$:
it seems clear that, for that value of $q$, the asymptotic regime has not yet been reached, for the lattice sizes
we simulated (in fact, it  is not clear if there is an asymptotic regime) \cite{girardi}. 
We would like to note that a good indication of the result quality is the relation $<n> \propto L^{\gamma/\nu}$. 
As we can see in Figs. 
\ref{fig:wolff}$b$ and  \ref{fig:wolff}$d$, our estimates of $\gamma/\nu$ from a log-log plot of
$<n>$ vs. $L$ (namely, $\gamma/\nu = 1.734 \pm 0.001$ and $1.7498 \pm 0.0009$ for $q=3$ and $4$,
respectively)
are, within error bars, the same of the (conjectured) exact results (namely, $\gamma/\nu=26/15=1.73333....$
for $q=3$ \cite{wu} and $\gamma/\nu=7/4=1.75$ for $q=4$ \cite{wu,creswick}).
		
	In order to compare our results for different values of $E_0$, we have plotted $z$ vs. $E_0$ for both $q=3$ and 
$q=4$ (see Fig. \ref{fig:zxE0}): we see that $z$ is approximately constant for any $E_0 \neq 0$ and strongly decreases
for $E_0=0$. In fact, the subtle decrease of $z$ as $E_0$ approaches $0$ may be a crossover effect: the closer
we get to the Wolff algorithm, the greater the value of $\tilde{L}$ and one has to go to larger and larger lattices
to obtain the correct dynamic behavior.
	
	Therefore, we have shown that the Wolff algorithm is still the most efficient procedure, when compared to
the generalizations for $E_0<0$. But we still have to check the dynamic behavior for $E_0>0$.
We know that for $E_0 \rightarrow \infty$ 
the algorithm is not ergodic. So we expect that, if a better value for $E_0$ exists, when compared to $E_0=0$,
it is not much greater than this last value. To address this
question we simulate the cases $E_0=0.05$ and $0.1$. 
In Fig. \ref{fig:grande} we show the results for $q=3$ (a) 
and $q=4$ (b), both calculated from the energy autocorrelation function. 
As we can see, for $q=3$ the autocorrelation time $\tau$ for $E_0=0.05$ is much greater than for 
$E_0=0$ and grows faster than for the latter. For $q=4$, $\tau$ is slightly 
greater for $E_0=0.05$ than for $E_0=0$, although our result is consistent with the same value of $z$ for both cases. However,
one has to consider that the implementation of the algorithm for $E_0=0.05$ is more complex than for $E_0=0$.

\section{Conclusion} \label{sec:summary}

       In this work we studied the Niedermayer algorithm applied to the two-dimensional Potts model with 2, 3, and 
 4 states. Our goal was 
 to determine which value of $E_0$ leads to the optimal algorithm, i.e.,
 to the smallest value of the dynamic exponent $z$. We observe that for $-1 \leq E_0 < 0$ there is a lattice size $\tilde{L}$,
 such that, for $L \geq \tilde{L}$, the average size of updates clusters is constant and the dynamic behavior of the algorithm is the same as
 for Metropolis'. The value of $\tilde{L}$ increases with  $E_0$ and diverges for
 $E_0=0$ (Wolff algorithm). When we look to the auto-correlation function, we notice that for $L < \tilde{L}$ the auto-correlation 
 time of the energy is greater than for the magnetization and the opposite happens for $L \geq \tilde{L}$. As we show in Fig. \ref{fig:magnatizacaoeenergia}, the quantity with greater autocorrelation time have the auto-correlation function 
 well described by a single exponential. For $E_0=0$ we regain the Wolff algorithm. 
 
    In Table 1 we summarize our findings,
 which show that the Wolff algorithm, $E_0=0$, is more efficient than its generalization for $E_0<0$. 
 There is still the possibility that some value of $E_0> 0$ may present a lower value of $z$, when
 compared to $E_0=0$. We show that, if this value exists, it is lower than $E_0<0.05$ and the complexity 
 of the algorithm is greater than the improvement in the dynamic behavior.

\begin{table}[h]
	\begin{center}
	    	\begin{tabular}{|c|c|c|} \hline
				q$\;$  & $-1\leq E_0 <0$ &Wolff ($E_0=0$)\\ \hline
				2  & 2.16(1) & undefined\\ \hline
				3  & 2.162(7)  & 0.55(2) \\\hline 
				4  & 2.21(2)  & 1.00(2) \\\hline 
	        \end{tabular}
	    \end{center}
	    \label{tab:resultado}
	    \caption{ Values for the dynamic exponent $z$ for the three models studied here and for $-1 \le E_0 \leq 0$.}
\end{table}
   
\section*{Acknowledgments}
The authors would like to thank the Brazilian agencies
FAPESC, CNPq, and CAPES for partial financial support.

\bibliographystyle{elsarticle-num}
\bibliography{references}

\begin{thebibliography}{10}
\expandafter\ifx\csname url\endcsname\relax
  \def\url#1{\texttt{#1}}\fi
\expandafter\ifx\csname urlprefix\endcsname\relax\def\urlprefix{URL }\fi
\expandafter\ifx\csname href\endcsname\relax
  \def\href#1#2{#2} \def\path#1{#1}\fi

\bibitem{landau}
D.~P. Landau, K.~Binder, A Guide to Monte Carlo Simulations in Statistical
  Physics, Cambridge University Press, New York, USA, 2000.

\bibitem{newman}
M.~Newman, G.~Barkema, {Monte Carlo Methods in Statistical Physics}, Oxford
  University Press, USA, 1999.

\bibitem{wang}
F.~G. Wang, D.~P. Landau, Phys. Rev. Lett. 86 (2001) 2050.

\bibitem{penna1990}
T.~Penna, P.~de~Oliveira, J. of Stat. Phys. 61 (1990) 933.

\bibitem{swendsen}
R.~H. Swendsen, J.-S. Wang, Phys. Rev. Lett. 58 (1987) 86.

\bibitem{wolff}
U.~Wolff, Phys. Rev. Lett. 62 (1989) 361.

\bibitem{metropolis}
N.~Metropolis, A.~Rosenbluth, M.~Rosenbluth, A.~Teller, E.~Teller, The Journal
  of Chemical Physics 21 (1953) 1087.

\bibitem{nightingale}
M.~P. Nightingale, H.~W. Blöte, Phys. Rev. Lett. 76 (1996) 4548.

\bibitem{baillie}
C.~Baillie, P.~Coddington, Phys. Rev. B 43~(13) (1991) 10617.

\bibitem{coddington}
P.~D. Coddington, C.~F. Baillie, Phys. Rev. Lett. 68 (1992) 962.

\bibitem{girardi}
D.~Girardi, N.~S. Branco, Journal of Statistical Mechanics: Theory and
  Experiment 2010 (2010) P04012.

\bibitem{niedermayer}
F.~Niedermayer, Phys. Rev. Lett. 61 (1988) 2026.

\bibitem{wu}
F.~Y. Wu, Rev. Mod. Phys. 54 (1982) 235.

\bibitem{wansleben}
S.~Wansleben, D.~P. Landau, Phys. Rev. B 43 (1991) 6006.

\bibitem{salas}
J.~Salas, A.~D. Sokal, J. of Stat. Phys. 87 (1997) 1.

\bibitem{binder}
K.~Binder, J. of Stat. Phys. 24 (1981) 69--86.

\bibitem{girardi1}
D.~Girardi, N.~S. Branco, Phys. Rev. E 83 (2011) 061127.

\bibitem{picco}
M.~Picco, Arxiv preprint cond-mat/9802092.

\bibitem{coddington2}
P.~Coddington, C.~Baillie, Nuclear Physics B-Proceedings Supplements 26 (1992)
  632.

\bibitem{li}
X.~Li, A.~Sokal, Phys. Rev. Lett. 63 (1989) 827.

\bibitem{creswick}
R.~J. Creswick, S.~Y. Kim, J. Phys. A: Math. Gen. 30 (1997) 8785.

\end{thebibliography}

\begin{figure}
	\begin{center}
	\leavevmode
	\caption{Magnetization and energy autocorrelation functions versus time (in $MCS$)
	for the Niedermayer algorithm with $E_0=-0.25$  and $q=3$. The main window represents the behavior for
	linear size $L=16$, while the inset applies to $L=128$.}
	\label{fig:magnatizacaoeenergia} %fig1
	\end{center}
\end{figure}

\begin{figure}
	\begin{center}
	\leavevmode
	\caption{Log-log graph  of magnetization autocorrelation time
	(in $MCS$) versus linear size $L$ for the Metropolis algorithm with $q=2$, $3$ and $4$. The quoted value for $z$ is obtained from the slope of a fitted straight line for the magnetization autocorrelation time.}
	\label{fig:metropolis} %fig2
	\end{center}
\end{figure}

\begin{figure}
		\begin{center}
		\leavevmode
		\caption{ Log-log graphs of magnetization (circle) and energy (square) autocorrelation times (in $MCS$) versus linear size $L$ for the Niedermayer algorithm with $E_0=-0.75$ for a) $q=3$ and c) $q=4$. Log-log graph of average cluster size $<n>$ versus $L$ for b) $q=3$ and d) $q=4$.}
		\label{fig:tao-0,75} %fig3
		\end{center}
	\end{figure}

\begin{figure}
		\begin{center}
		\leavevmode
		\caption{ a) Log-log graphs of magnetization (circle) and energy (square) autocorrelation time
		(in $MCS$) versus linear size $L$ for the Niedermayer algorithm with $E_0=-0.25$ for $q=3$ . b) Log-log graph of average cluster size $<n>$ versus $L$ for $q=3$.}
		\label{fig:tao-0,25} %fig4
		\end{center}
	\end{figure}

\begin{figure}
		\begin{center}
		\leavevmode
		\caption{ Log-log graphs of magnetization (circle) and energy (square) autocorrelation time
		(in $MCS$) versus linear size $L$ for the Niedermayer algorithm with $E_0=0.0$ (Wolff
		algorithm) for a) $q=3$ and c) $q=4$. Note that the values we quote for the dynamic exponent $z$ are those obtained from the greatest three values of $L$ (see text). Log-log graph of average cluster size $<n>$ versus $L$ for 
		b) $q=3$ and d) $q=4$.}
		\label{fig:wolff} %fig5
		\end{center}
	\end{figure}

	\begin{figure}
		\begin{center}
			\leavevmode
			\caption{ Semi-log graphs of $z$ versus $1/L_{min}$ (see the text) for $q=2$, 	$3$ and $4$.}
			\label{fig:picco} %fig6
		\end{center}
	\end{figure}

\begin{figure}
	\begin{center}
	\leavevmode
	\caption{Dependence of the dynamic exponent $z$ on $E_0$, for $-1 \leq E_0 \leq 0$
	and $q=3$ (circles) and $q=4$ (squares).}
	\label{fig:zxE0} %fig7
	\end{center}
\end{figure}

	\begin{figure}
			\begin{center}
			\leavevmode
			\caption{ Log-log graphs of energy autocorrelation time
			(in $MCS$) versus linear size $L$ for the Niedermayer algorithm with $E_0=0$ (circle) and $E_0=0.05$ (square) for a) $q=3$ and b) $q=4$. }
			\label{fig:grande} %fig8
			\end{center}
		\end{figure}
\end{document}